\documentclass[preprint,10pt]{aastex}
\usepackage[utf8]{inputenc}

\newcommand{\gaia}{{\it Gaia\/}}

\title{The canonical Luminous Blue Variable AG~Car and its neighbor Hen 3-519 are much closer than previously assumed}
\author{Nathan Smith\altaffilmark{1} \& Keivan G.\ Stassun\altaffilmark{2,3}}
\altaffiltext{1}{Steward Observatory, University of Arizona, 933 North Cherry Avenue, Tucson, AZ 85721, USA}
\altaffiltext{2}{Vanderbilt University, Department of Physics \& Astronomy, 6301 Stevenson Center Ln., Nashville, TN  37235, USA} 
\altaffiltext{3}{Fisk University, Department of Physics, 1000 17th Ave. N., Nashville, TN  37208, USA}

\begin{document}

\begin{abstract}
The strong mass loss of Luminous Blue Variables (LBVs) is thought to play a critical role in massive-star evolution, but 
their place in the evolutionary sequence remains debated. A key to understanding their peculiar instability is their high observed luminosities, which 
often depends on uncertain distances. Here we report direct distances and space motions of four canonical Milky Way LBVs---AG~Car, HR~Car, HD~168607, and (candidate) Hen~3-519---
from the \gaia\ first data release. Whereas the distances of HR~Car and HD~168607 are consistent with previous literature estimates within the considerable uncertainties, 
Hen~3-519 and AG~Car, 
both at $\sim$2~kpc, are much closer than the 6--8~kpc distances previously assumed. 
As a result, Hen~3-519 moves far from the locus of LBVs on the HR Diagram, making it a much less luminous object. For AG~Car, considered a defining example of a classical LBV, its lower luminosity would also move it off the S~Dor instability strip. Lower luminosities allow both AG~Car and Hen~3-519 to have passed through a previous red supergiant phase, lower the mass estimates for their shell nebulae, and imply that binary evolution is needed to account for their peculiarities. These results may also impact our understanding of LBVs as potential supernova progenitors and their isolated environments. Improved distances will be provided in the \gaia\ second data release, which will include additional LBVs. AG~Car and Hen~3-519 hint that this new information may alter our traditional view of LBVs.
\end{abstract}

\section{Introduction\label{sec:intro}}

Eruptive mass loss exhibited by luminous blue variables (LBVs) is
thought to be important for the evolution of massive stars
\citep{hd94,so06,smith14}, but the exact role LBVs play and the physics of their instability has been challenging to understand.  The 
class of LBVs defined in the Milky Way and Magellanic Clouds (LMC/SMC)
is thought to be responsible for some extragalactic non-supernova (SN)
transients \citep{smith+11,vdm12}.  The episodic mass loss of LBVs has
also been a reference point for interpreting the dense circumstellar
material (CSM) around Type~IIn supernovae (SNe~IIn), which is thought
to be indicative of extreme pre-SN eruptions \citep{smith14}.

The diverse collection of objects known collectively as LBVs was
first proposed as a group by \citet{conti84}, and the now standard
interpretation of these stars and their role in evolution was
established through the 1980s and 1990s.  The review by \citet{hd94}
provides a summary of the traditional view of LBVs, which has a few
key tenets:

1.  LBVs are thought to represent a brief transitional phase in the
evolution of the most massive stars, between the main sequence O-type
stars and the H-deficient Wolf-Rayet (WR) stars.  Because they have
already lost some mass and because their core luminosity has
increased, the $L/M$ ratio is high.  This proximity to the classical Eddington limit, combined with Fe opacity, leads to a violent instability in the envelope that
somehow triggers runaway Geyser-like mass loss.  The heavy mass loss
of LBVs, in the single-star view, is essential to remove the H
envelope to form WR stars.

2.  Observed S~Doradus-type variations, as exemplified most clearly by
AG~Car in the Milky Way and R127 in the LMC \citep{ws82,stahl83}, are
temperature variations at constant bolometric luminosity.  The cool
(visibly bright) state is caused by a pseudo photosphere that develops
in an optically thick wind, which occurs at the same temperature
regardless of luminosity \citep{davidson87,hd94}.  In their quiescent
(hot) phase, all LBVs reside on the diagonal ``S~Doradus instability
strip'' \citep{wolf89} on the Hertzsprung-Russell (HR) Diagram.

3.  The strong mass loss of LBVs halts their redward evolution,
preventing them from becoming red supergiants (RSGs).  This explains
the observed absence of high-luminosity RSGs above
log($L/L_{\odot}$)=5.8, due to instability and mass loss in evolved
massive single stars.

4. Some LBVs suffer dramatic ``giant eruptions'' where huge amounts of
mass can be lost.  Although there is a formalism to explain the
mechanism of this mass loss with super-Eddington continuum-driven
winds \citep{owocki04}, the underlying reason why single stars
suddenly exceed the classical Eddington limit by factors of 5-10 has
remained unexplained.  Nevertheless, this tremendous mass loss is
observed and is likely to be important in the evolution of massive
stars.

Several aspects of this traditional view, however, have unraveled 
with time.  In particular, the important conjecture that S~Dor
variations are major mass-loss events caused by optically thick winds
appears to be wrong, and steady super-Eddington winds might not be the
dominant driving mechanism in many giant LBV eruptions (see review by
\citealt{smith14}).  Morover, the recognition that SNe~IIn have
progenitors consistent with LBVs (see, e.g., \citealt{mauerhan13})
casts doubt on the idea that LBVs are only a brief transitional phase
before the start of core He burning.  Investigations with quantitative
spectroscopy \citep{dekoter96,groh09a,groh09b} disproved the
conjecture that S~Dor brightening events are caused by
pseudo-photospheres in optically thick winds (\citealt{davidson87}).
The mass-loss rates of S~Dor maxima are not high enough to make such
extreme pseudo-photospheres, and so they are more likely to be caused
by envelope inflation or pulsation \citep{gov12}.  Also, bolometric
luminosities during S~Dor eruptions are not necessarily constant
\citep{groh09a}.  Similarly, the idea that giant-eruption maxima are
caused by pseudo-photospheres in super-Eddington winds is challenged
by light-echo spectra of $\eta$~Carinae \citep{rest12,prieto14}
(although see \citealt{owocki16}), by detailed analysis of the ejecta
around $\eta$ Car that are better matched by an explosive event
\citep{smith06,smith08,smith13}, and the fact that many extragalactic
giant LBV eruptions are relatively hot at peak luminosity rather than
cool \citep{smith+11,mauerhan15}.  Last, as discussed by \citet{st15},
their isolation from massive O-type stars suggests that they are much
older than expected.  \citet{st15} suggested that they are
largely products of binary evolution and not a transitional state in
the lives of the most massive single stars.

Here we add another possible wrinkle to the unraveling story of LBVs.  One of
the canonical, classical high-luminosity LBVs in the Milky Way is
AG~Carinae; from a direct measurement of its parallax from \gaia\,
this star appears to be much closer -- and therefore much less
luminous --- than previously thought.  Even though AG~Car is a
defining S Doradus variable, this closer distance moves it well below
the S Doradus instability strip established for LBVs in the LMC
\citep{wolf89}.  While the distance to LBVs in the LMC is not so
uncertain, the distance to Galactic LBVs has always been precarious.
Below we discuss AG~Car and the three other Galactic LBV-like stars
that are included in the first \gaia\ data release.

The \gaia\ first data release (DR1) has recently provided trigonometric parallaxes and proper motions for the $\approx$2 million stars previously observed by {\it Hipparcos\/} and {\it Tycho-2\/}, with a typical precision in the parallax of $\approx$0.25~mas \citep{Gaia:2016}.  These parallaxes are enabling a number of investigations not previously possible, such as direct measurement of the basic properties of all planets and their stellar hosts \citep{Stassun:2016}.  These fundamental new measurements also provide an opportunity to reassess the nature of these LBVs. 

The data from \gaia\ DR1 and from the literature that we use in this study are described in Section~\ref{sec:data}, which in particular includes parallaxes and proper motions for the four Milky Way LBVs Hen~3-519, HD~168607, HR~Car, AG~Car. Section~\ref{sec:results} presents the principal results of this study, namely, the distances, space motions, and other basic characteristics of these LBVs, as well as the uncertainty in these values.  Section 4 discusses the implications of our findings, especially the distances for AG~Car and Hen~3-519 which we find to be much closer than previously thought.  We conclude with a brief summary of our results in Section~\ref{sec:summary}.

\section{Data and Methods\label{sec:data}}

We searched the \gaia\ DR1 catalog for all of the Milky Way LBVs and LBV candidates listed in \citet{st15}, which yielded parallaxes and proper motions for four: Hen~3-519, HD~168607, HR~Car, AG~Car (see Table~\ref{tab:data}). From the parallax, $\pi$, a distance may be straightforwardly computed via $d = 1/\pi$.  The parallaxes of the four LBVs span the range 0.34--1.02~mas, corresponding to distances spanning the range $\approx$1--3~kpc.

\begin{deluxetable}{lccrrrrrrrrrr}
\tabletypesize{\scriptsize}
\rotate
\tablewidth{0pt}
\tablecolumns{13}
\tablecaption{Data used for the LBVs in our study sample\label{tab:data}}
\tablehead{
\colhead{Name} & \colhead{Hipp. ID} & \colhead{TYC ID} & \multicolumn{6}{c}{{\it Gaia} DR1} & \multicolumn{4}{c}{\citet{Astraatmadja:2016}} \\ 
\colhead{} & \colhead{} & \colhead{} & \colhead{$\pi$ (mas)} & \colhead{$\sigma_\pi$ (mas)} & \colhead{$\mu_\alpha$ (mas yr$^{-1}$)} & \colhead{$\sigma_{\mu_\alpha}$ (mas yr$^{-1}$)} & \colhead{$\mu_\delta$ (mas yr$^{-1}$)} & \colhead{$\sigma_{\mu_\delta}$ (mas yr$^{-1}$)} & \colhead{$d$ (pc)} & \colhead{$\sigma_d$ (pc)} & \colhead{$d_{5\%}$ (pc)} & \colhead{$d_{95\%}$ (pc)}
}
\startdata
Hen 3-519 & \nodata & 8958-1166-1 & 0.796 & 0.575 & $-$4.282 & 1.799 & 3.904 & 0.853 & 1600 & 859 & 750 & 3575 \\ 
HD 168607 & 89956 & \nodata & 1.016 & 0.282 & 0.404 & 0.102 & $-$1.314 & 0.063 & 1156 & 346 & 754 & 1891 \\ 
HR Car & 50843 & \nodata & 0.342 & 0.239 & $-$5.813 & 0.069 & 2.902 & 0.065 & 2267 & 973 & 1407 & 4607 \\
AG Car & 53461 & \nodata & 0.400 & 0.225 & $-$4.861 & 0.079 & 1.923 & 0.092 & 1953 & 726 & 1317 & 3705 \\
\enddata
\end{deluxetable} 

However, because the parallax measurement errors are relatively large (28\% to 72\%), the distances and their uncertainties are expected to depend on the adopted prior \citep[see, e.g.,][]{Bailer-Jones:2015}.  Therefore, we have also retrieved the distances computed by \citet{Astraatmadja:2016} using a more realistic prior distribution based on simple but empirically motivated stellar density distributions for the Milky Way. These distances, which span the range 1.2--2.3~kpc (Table~\ref{tab:data}), are consistent within the uncertainties with those obtained simply via $1/\pi$, but with somewhat smaller uncertainties due to the more informative prior used. Thus we prefer the \citet{Astraatmadja:2016} distances (hereafter, ABJ) in what follows but note that our primary conclusions do not depend strongly on this choice. 

The \gaia\ DR1 release notes state that the parallaxes may possess uncharacterized systematic uncertainties of up to 0.3~mas, and recommend adding an additional 0.3~mas to the reported measurement uncertainty. Consequently, \citet{Astraatmadja:2016} provide a version of their catalog that includes this additional 0.3~mas error. However, \citet{Stassun:2016b} used a set of nearby, benchmark eclipsing binary stars \citep{Stassun:2016a} to quantify the systematic error, finding it to be $-0.25\pm 0.05$~mas in the sense that the \gaia\ parallaxes are systematically too small (distances too long). 
\citet{Jao:2016} found a similar offset among a sample of nearby, high-proper-motion stars in the solar neighborhood. 
However, at larger distances, \citet{Stassun:2016a} found that the systematic offset vanishes for $\pi\lesssim 1$~mas ($d \gtrsim 1$~kpc). 
This is corroborated by \citet{Casertano:2016}, who found excellent agreement between the \gaia\ distances and a large sample of Galactic Cepheids at $d \sim 2$~kpc.
\citet{Lindegren:2016} and \citet{Sesar:2016} similarly use samples such as RR Lyrae stars at intermediate distances to argue that no correction is needed beyond $\sim$1~kpc. 
Finally, \citet{Davies:2017} uses red clump stars over a large range of galactic distances to broadly confirm the above findings, with an offset similar to that of \citet{Stassun:2016a} for nearby stars that vanishes at $d \gtrsim 1.2$~kpc. 
Therefore, since the distances of the LBVs in our study sample are all greater than $\sim$1~kpc (Table~\ref{tab:data}), we do not apply any offset to the parallaxes and moreover do not add any further systematic error to the reported measurement uncertainties.

\section{Results\label{sec:results}}

{\it HD~168607:} This LBV is located only about 1{\arcmin} away from HD~168625, which is well-known for its SN~1987A-like triple-ring nebula \citep{smith07}.  HD~168607 is considered an LBV based on its characteristic variability, whereas its neighbor HD~168625 is usually
considered an LBV candidate \citep{sterken99}.  Both are found in the
outskirts of the star-forming region M17, and are thought to be part
of the larger Ser OB1 association at $\sim$2.2 kpc \citep{cg04}.  A
distance of $\sim$2.2 kpc is usually adopted in the literature (van
Genderen et al.\ 1992), although distances of 1.2--2.8 kpc have been
proposed \citep{rh98,pasquali02}.  The \gaia\ DR1 distance to HD~168607
from Table~1 is 0.98$\pm$0.27~kpc ($\pi = 1.02\pm 0.28$ mas) or
1.16$\pm$0.35~kpc ($\pi = 0.87\pm 0.26$ mas) using the ABJ parallaxes. This is on the low end of
previously adopted values that are usually close to 2 kpc, but not too
far off.  The 95\% upper limit from the ABJ parallaxes is 1.9 kpc.  This is unlikely to prompt
a major revision in the interpretation of this object.  We postpone a
detailed discussion until the higher precison that will be available
in the next \gaia\ data release.  HD~168607 appears to have a low
transverse velocity of only a few km s$^{-1}$ indicated by its \gaia\
proper motion.

{\it HR Car:} The distance adopted in the literature for HR~Car is
usually 5$\pm$1~kpc (e.g., \citealt{vangenderen91,groh09a}), placing
it among the low-luminosity group of LBVs (see, e.g., \citealt{svdk04}).
The \gaia\ distance in Table~1 is 2.92$\pm$2.04~kpc ($\pi = 0.34\pm 0.24$ mas) or
2.27$\pm$0.97 kpc ($\pi = 0.44\pm 0.19$ mas) according to the ABJ parallaxes, which is
lower than previously assumed, but the 95\% upper limit of 4.6~kpc is arguably consistent with the usual estimate.
Given that HR~Car is also now known to have a resolved wide companion
about 3~mas away \citep{boffin16}, which may complicate the measured
parallax, we do not advocate a major revision of its distance or
luminosity at this time.  Again, we postpone a detailed discussion
pending the higher precision that will be available in the next \gaia\
data release.  
The \gaia\ absolute proper motion of HR Car is about
6.5$\pm$0.16~mas~yr$^{-1}$ (PA$\approx$297{\arcdeg}), which at a
distance of $\sim$2.5~kpc, translates to a transverse velocity of
about 72~km~s$^{-1}$.

{\it AG Car:} Although AG~Car is seen in projection amid the Car
OB1/OB2 association, located at 2--2.5~kpc, a larger distance has
usually been adopted in the literature, making AG~Car one of the most
luminous stars in the Milky Way.  The larger distance is
based on its radial velocity relative to the local standard of rest
(LSR) as compared to the Galactic rotation curve, as well as its high
line-of-sight extinction, from which \citet{humphreys89} derived a
likely distance of 6.4--6.9~kpc.  However, the new \gaia\ distance is
2.50$\pm$1.41~kpc ($\pi = 0.40 \pm 0.23$ mas) or
1.95$\pm$0.73~kpc ($\pi = 0.51 \pm 0.19$ mas) according to the ABJ parallaxes, with a 95\% upper limit of 3.7~kpc (Table~1).  This
would appear to rule out the larger distance above 6 kpc derived by
\citet{humphreys89}, instead suggesting membership in the closer Car
OB1/OB2 association after all.  If even approximately correct, this much closer distance has profound implications for our interpretation of AG~Car, and consequences for
our understanding of LBVs in general, as discussed in the next
section.  At a distance of $\sim$2~kpc, the measured \gaia\ absolute proper
motion of 5.2$\pm$0.26~mas~yr$^{-1}$ (to the west/northwest;
PA$\approx$292{\arcdeg}), would suggest a transverse
velocity of about 50~km~s$^{-1}$.

{\it Hen 3-519:} Hen~3-519 is located very close on the sky to AG~Car
(about 20{\arcmin} away) and would appear to be part of the same Car
OB1/OB2 association, but like AG~Car, previous authors have generally
favored a very large distance near 8~kpc.  The large 8~kpc distance
was proposed by \citet{davidson93} based on the large line-of-sight
extinction inferred from UV data.  \citet{smith94} found a slightly
lower reddening, but also favored a large distance near 8 kpc based on
the LSR velocity implied by nebular emission and interstellar
absorption lines in high-resolution spectra.  Again, this would make
Hen~3-519 an extremely luminous star.  Like AG~Car, though, parallax
once again indicates a much smaller distance.  The \gaia\ distance is
1.26$\pm$0.91~kpc ($\pi = 0.80 \pm 0.58$ mas) or 
1.60$\pm$0.86~kpc ($\pi = 0.63 \pm 0.34$ mas) according to the ABJ parallaxes, with a 95\% upper limit from the ABJ parallaxes of 3.6~kpc (Table~1).  This
would appear to rule out the larger distance of 8~kpc proposed by
\citet{davidson93}, instead suggesting membership in the closer Car
OB1/OB2 association, just like its neighbor AG~Car.  At a distance of
$\sim$2 kpc, the measured \gaia\ absolute proper motion of 5.8$\pm$2.7 mas
yr$^{-1}$ (PA$\approx$312{\arcdeg}), would suggest a
transverse velocity of 56~km~s$^{-1}$.  This is, again, very similar
in magnitude and direction to the motion of AG Car, although with
larger uncertainty.

The three targets in Carina all exhibit a similar absolute proper motion of roughly 5~mas~yr$^{-1}$ to the west/northwest.  This corresponds to apparent motion in a direction along the Galactic plane.  It is roughly consistent with the direction and magnitude of longitude drift expected for Galactic rotation on circular orbits in the Solar neighborhood.  At $\sim$2~kpc, objects on the near side of the Carina Arm are at roughly the same radius from the Galactic Center as the Sun.  At a much larger distance of 6--8 kpc and a location on the far side of the Carina Arm, the expected proper motions should be less, so perhaps the improved precision of future \gaia\ data releases will provide a more definitive constraint from proper motion.

\begin{figure}[!ht]
    \centering
    \includegraphics[width=0.8\linewidth,trim=10 80 10 80,clip]{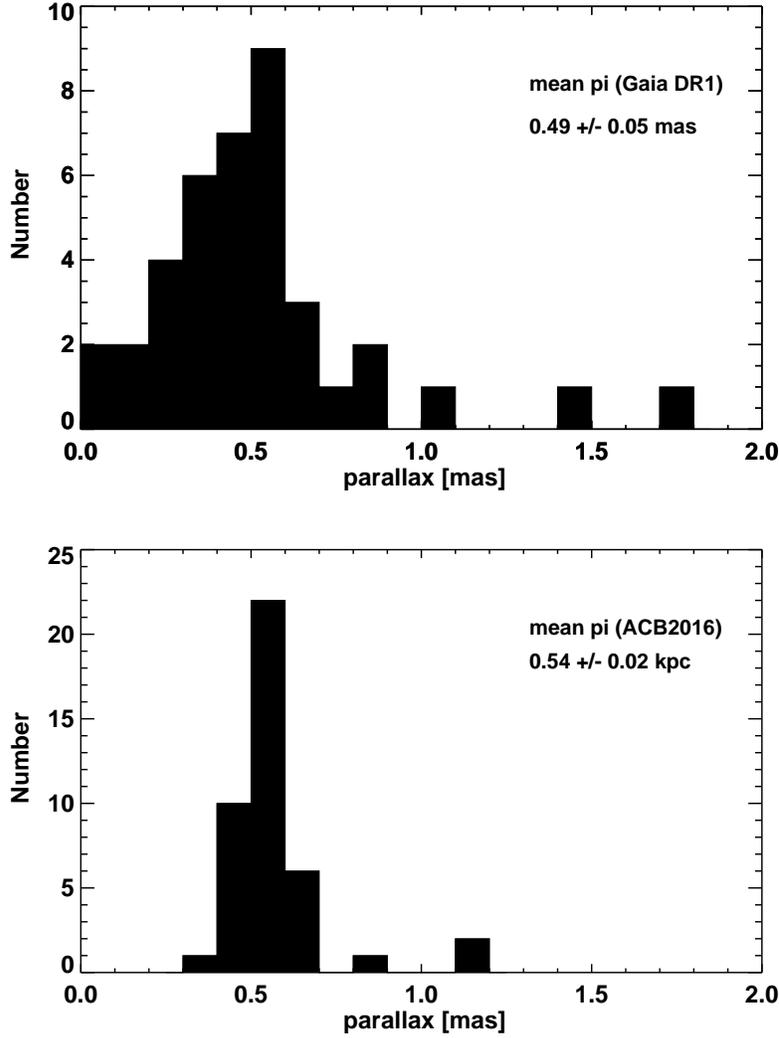}
    \caption{(Top:) Distribution of \gaia\ DR1 parallaxes for O stars in the Carina star-forming region from \citet{Smith:2006}, all expected to be at $\pi\sim$0.45~mas ($d\sim 2.2$~kpc). 
    (Bottom:) Parallaxes according to a Milky Way prior based calculation \citep{Astraatmadja:2016}.}
    \label{fig:smith06}
\end{figure}

As a further check on the veracity of the \gaia\ distances for these stars, we used the sample of O stars in the Carina star-forming region\footnote{Note that we are not making this comparison to suggest that AG~Car or Hen~3-519 are necessarily members of the Carina OB association (although they might be), and we do not depend upon this assumption.  Instead, we are using the Carina Nebula O-type stars as an example of a cluster in the same region of the sky where the distance is known reliably from other information, as a test case to check the validity of the DR1 results, in particular for luminous stars. The distance we derive from DR1 is roughly correct, and the errors are similar to our LBV targets discussed in this paper.} from \citet{Smith:2006}, which are all expected to be at a common distance of $\sim$2.2~kpc ($\pi\sim 0.45$~mas) on the basis of multiple lines of evidence, including the nebular expansion of $\eta$~Car itself \citep[see, e.g.,][]{smith06}. Forty-three of the stars in that study are present in the \gaia\ DR1, and in Figure~\ref{fig:smith06} we present the distribution of their parallaxes using both the 
direct \gaia\ DR1 parallaxes and the Milky Way prior based parallaxes from \citet{Astraatmadja:2016}. The former gives a mean distance of 2.06$\pm$0.41~kpc and the latter gives 1.86$\pm$0.05~kpc. The median \gaia\ parallax error for these stars is 63\%, comparable to that for the four LBVs in our study sample.

Thus, the \gaia\ distances for the Carina O stars via a simple $1/\pi$ estimator are consistent with expectation; the distances from the Milky Way prior based method \citep{Astraatmadja:2016} are perhaps slightly underestimated. 
In any event, this check provides a measure of validation that the \gaia\ distances for our target LBVs---also luminous stars at expected distances $\sim$2~kpc or greater---should be reliable \citep[see also][]{Stassun:2016b,Casertano:2016,Jao:2016,Davies:2017}. 
If so, the implications are important, as discussed next.

\begin{figure*}\begin{center}
\includegraphics[width=5.0in,trim=0 0 100 250,clip]{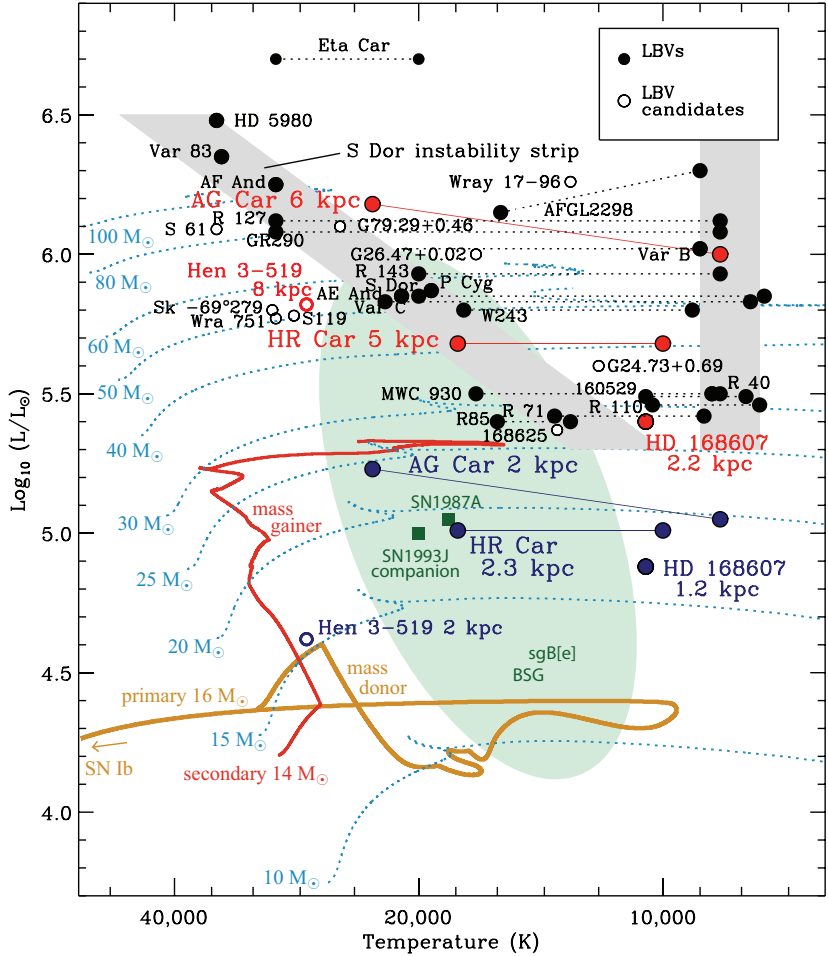}
\end{center}
\caption{The HR Diagram of LBVs.  This is the same as in \citet{st15},
  except that we have modified the entries for AG Car, HR Car,
  HD~168607, and Hen 3-519 as discussed in this paper.  For all three,
  values from the literature with conventional distances are shown in
  red, and these are reduced according to the lower distances
  indicated by new \gaia\ parallax measurements (shown in blue).  For
  AG Car, we use the values derived from CMFGEN models assuming $d$=6
  kpc from \citet{groh09b,groh11} in red, and the same temperatures
  but at $d$=2 kpc in blue. For HR Car we use values derived from
  CMFGEN models by \citet{groh09a} at 5 kpc (red) and 2.3 kpc (blue).
  For Hen 3-519 we use the temperature derived from CMFGEN models by
  \citet{smith94}.  We adopt the average of two luminosities from
  \citet{smith94} and \citet{davidson93} at 8 kpc (red), and at 2 kpc
  (blue).  For HD~168607, we use the values from \citet{lw84}.  As in
  the original figure of \citet{st15}, single-star evolutionary tracks
  are from \citet{brott11}, and example binary tracks are from
  \citet{lk14}.}
  \label{fig:hrd}
\end{figure*}

\section{Discussion\label{sec:disc}}

Figure~\ref{fig:hrd} shows LBVs on the HR Diagram, adapted from
previous studies as noted in the caption.  The red points show the
luminosities for AG~Car, HR~Car, HD~168607, and Hen~3-519 using
previously adopted distances in the literature, whereas the blue
points show how the luminosities are lowered if the nearer
\gaia\ DR1 distances are adopted.  For Hen~3-519, widely different
luminosities are given for the same 8 kpc distance by
\citet{davidson93} and \citet{smith94} based on different assumed
values of $E(B-V)$.  We adopt an average of the two here, although
this difference is small compared to the factor of 16 reduction in
luminosity that results from the nearer distance.  

As noted above, the nearer distance for HR~Car is not in such severe
disagreement with previous values, and for HD~168507, the distance
agrees within the quoted uncertainty with previous estimates.  It is
interesting, though, that at the nearer 2.3 kpc distance, the hotter
state of HR Car is almost coincident on the HR Diagram with the
progenitor of SN~1987A.  Speculative comparisons between SN~1987A's
progenitor and LBVs have been discussed before \citep{smith07}.
Future \gaia\ data releases will provide higher precision in the
parallax and proper motion, so we postpone a more detailed discussion
of HR~Car and HD~168607 until then.  For now, we discuss the basic
implications of the substantially smaller distances for AG~Car and
Hen~3-519, where the \gaia\ distances---even with the relatively large uncertainty of DR1---appear to be inconsistent with previous
estimates, and where the possible implications for our understanding
of LBVs are more severe.  We acknowledge that the following discussion may be need to be revisited if the improved \gaia\ DR2 measurements show the adopted DR1 values and uncertainties to be substantially incorrect.

\subsection{AG Car as a  prototypical S Doradus variable} 

The new distance of $\sim$2 kpc implied by AG~Car's parallax is over 3
times closer than the usually assumed value.  \citet{humphreys89} preferred 6.4-6.9 kpc, but at that distance, the parallax would be only 0.16 mas or less.  Most authors have generally
adopted a value of 6 kpc, so the $\sim$3 times closer \gaia\ distance
makes AG~Car about 9 times less luminous than previously assumed.  If true, this would have a tremendous
impact on our interpretation of this star and possibly LBVs in
general, and we explore these implications below.  To place AG~Car on the HR Diagram, we assume that the
temperatures derived from CMFGEN models by \citet{groh09b,groh11} are
the same, but that the bolometric luminosity scales by the square of
the assumed distance.  There are several immediate implications:

1.  If AG Car's luminosity is lower by a factor of 9, it cannot be
such a massive star.  Previously it sat along an evolutionary track
for an initial mass of around 90-100 $M_{\odot}$, but the lower
luminosity corresponds to a single star of only $\sim$25 $M_{\odot}$
initial mass.  Single stars of this mass are not expected to approach
the classical Eddington limit during their evolution.  This raises
profound questions about the source of AG~Car's instability and mass
loss, and the nature of the LBV instability itself.

2.  Lowering its luminosity moves AG~Car far off the S Dor instability
strip, which has been seen as the defining locus of LBVs in their hot
state \citep{hd94}.  Yet, AG~Car is an S Dor variable; in fact, it is
a defining member of the class \citep{stahl86,ws82,stahl01}.  Since
AG~Car is considered a prototypical example of a classical
high-luminosity LBV \citep{ws82,stahl86,hd94}, this raises important
questions about our fundamental picture of LBVs, and suggests that the
S~Dor instability strip may not be as clean as previously thought -- at
least in the Milky Way.  Stars far away from the S~Dor instability
strip may evidently be prone to LBV-like behavior as well, if the
nearer \gaia\ distance is correct.

3. The new luminosity is only log($L/L_{\odot}$)=5.25, implying an initial mass (if single) of only $\sim$25 $M_{\odot}$, as noted above.  This is now low enough that AG~Car could have gone through a previous RSG phase. Several previous studies have concluded that dust in the shell nebula around AG~Car resembles dust seen around massive RSGs
\citep{smith97,voors00}.  The possibility of a previous RSG phase has
generally been discounted, since there are no RSGs observed at very
high luminosities corresponding to initial masses above 35-40
$M_{\odot}$, whereas AG~Car was thought to far exceed this limit.  The
new lower distance and $L$ removes this restriction.  With strong mass loss during the RSG phase, even single-star models can produce LBV-like stars in this lower 20-25 $M_{\odot}$ initial mass range \citep{groh13}.

4.  The lower distance also results in a lower mass for the nebula.
AG Car's shell nebula was thought to have a huge total mass of 20-25
$M_{\odot}$, derived from the measured dust mass of
$\sim$0.2-0.25 $M_{\odot}$ \citep{voors00,vn15}, an assumed gas:dust
mass ratio, and a distance of 6 kpc. Since the dust mass depends
linearly on the IR luminosity, the 3 times closer distance implies a 9
times lower nebular mass of roughly 2.5 $M_{\odot}$.  The time
averaged mass-loss rates needed to produce this nebula are therefore
not as extreme as previously thought.

5.  A closer distance for AG~Car also has important consequences
concerning the recently discussed environments of LBVs.  \citet{st15}
pointed out that in general, LBVs seem strangely isolated from other
young massive stars, and questioned their traditional role in
evolution as very massive single stars in transition from O-type stars to
Wolf-Rayet (WR) stars.  AG Car was surprisingly isolated for a star of
90-100 $M_{\odot}$ initial mass, being at least 30 pc in projection
from any other O-type star \citep{st15}.  Most of these neighboring
O-type stars were thought to be in the foreground at around 2 kpc,
while \citet{cg04} were unable to identify any host cluster or
association at $\sim$6 kpc that might trace AG~Car's birth environment.
Moving AG~Car to a distance of 2 kpc in the Car OB association
would mean that the projected separations on the sky to neighboring O-type
stars are relevant.  The large 30 pc separation would be
problematic if AG~Car was a 100 $M_{\odot}$ star --- but it is far less problematic if AG~Car has a much lower luminosity (and thus a lower initial mass and longer lifetime) than previously thought.\footnote{One of the most pressing mysteries of LBVs is their relative isolation as
  compared to other massive stars \citep{st15}.  This argument was
  based largely on the cumulative distribution of separations between
  stars in the LMC, so this is not affected by the smaller distance
  and luminosity of AG Car.}

So, with a nearer distance and lower luminosity, how can we understand
AG~Car's instability and mass loss that have been used to help define
the LBV class?  In the traditional picture, the combination of a very
high luminosity near the classical Eddington limit and rapid rotation
bring the star to the so-called ``$\Gamma\Omega$-Limit'' during its
post-main-sequence evolution (see \citealt{groh11} and references
therein). Indeed, AG~Car is thought to be a rapid rotator based on
evidence for a bipolar wind from spectropolarimetry, line profile
shapes, and other diagnostics of rotation \citep{sl94,claus94,groh06}.
However, this picture must be modified if AG~Car is really as close as
\gaia\ suggests.  Stars of much lower luminosity and an initial mass
around $\sim$25 $M_{\odot}$ might conceivably develop a high $L$/$M$
ratio if they shed a large amount of mass in a previous RSG phase, so
that similar ideas about envelope instability might still apply ---
{\it but it is hard to imagine that a single star could shed all that
  mass while maintaining rapid rotation}.  Instead, it seems as though
binary interaction must play some role to explain AG~Car, if it
is less massive than previously thought.  An example track of a
mass-gainer on the HR Diagram (Fig.~\ref{fig:hrd}; evolutionary track
from \citealt{lk14}) would seem to explain the current properties of
AG~Car quite naturally.  A stellar merger might
produce an equally satisfying explanation \citep{pod10}.  Such a star
gains angular momentum from the mass it accretes from a companion or
from a merger, rather than shedding all its angular momentum in the
RSG wind.

An important outstanding question is whether or not this rapid
rotation gained from binary interaction plays some critical role in
triggering the envelope instability that leads to S Doradus-like
variability.  \citet{gov12} have hypothesized that rotation
might be a trigger for the radius inflation of LBVs in their S Doradus
cycle, if this is brought on by the density inversion below the
photosphere that results from the Fe-opacity bump.  Based on their
surprisingly isolated environments (implying long lifetimes), \citet{st15} suggested that most LBVs may be products of mass accretion or mergers in interacting binaries.  This could potentially explain their anomolously high luminosities via rejuvenation, as well as rapid rotation late in life. If binary interaction is a key ingredient for the LBV instability,
this is not necessarily limited to only the most massive and luminous
stars.  

Overall, the possibility of moving AG~Car down on the HR Diagram raises intriguing new questions about the blue supergiants (BSGs) and B[e] supergiants
corresponding to initial masses of 15--40~$M_{\odot}$ (the green oval
in Fig.~\ref{fig:hrd}).  This region includes persistently mysterious
objects like the progenitor of SN~1987A, among others.  Finding
bona-fide LBVs like AG~Car and HR~Car in this same region may focus
our attention and generate new ideas about their significance in
stellar evolution.

\subsection{Hen 3-519 at low luminosity} 

Hen~3-519 has been studied far less intensively than AG~Car.  Less is
known about it, and it is not a critical defining LBV like AG~Car.
The luminosity at 8~kpc that we used in Figure~\ref{fig:hrd} is an
average of two values in the literature that disagree significantly
\citep{davidson93,smith94}.  We used an average of these to scale to the nearer distance of $\sim$2~kpc indicated by \gaia\ parallax.  Renewed efforts
to determine the fundamental physical parameters of this star are
encouraged.  The uncertainty in the measured parallax ($\sigma_{\pi}$ in Table~\ref{tab:data}) is larger than the other stars, so the precise value of this new distance should be viewed with some degree of caution.

It is admittedly quite surprising that a star of such low luminosity
would exhibit a peculiar emission-line spectrum that is so similar to
the spectra of much more luminous stars, such as the Ofpe/WN9 stars in
the LMC \citep{crowther95}.  Aside from the lack of any known
S~Dor-like variability in Hen~3-519, many of the other comments above
about AG~Car may apply here as well---especially regarding the role
of a previous RSG phase in forming its shell nebula, and the likely
importance of binary evolution in explaning its current location on
the HR diagram.  Perhaps Hen~3-519 is not related to the LBVs at all.
Its low luminosity might be more consistent with post-RSG evolution of
a 10--15~$M_{\odot}$ star, especially if it has encountered binary
interaction (i.e. post common-envelope or post-RLOF).  Regarding its LBV-like
spectrum, it may be relevant to point out that even some evolved 
intermediate-mass (5--8~$M_{\odot}$) stars inside bipolar planetary 
nebulae have spectra that very closely
resemble the unusual spectrum of the supermassive LBV $\eta$~Carinae
\citep{balick89,smith03}.  Evolved stars of vastly different luminosity and mass may achieve similar temperatures, densities, and composition in their winds.

\section{Summary and Conclusions\label{sec:summary}}

We report the parallax distances and proper motions of the Galactic
LBV stars AG~Car, HR~Car, and HD~168607, and the LBV candidate
Hen~3-519, resulting from the \gaia\ DR1 first data release. These are
the only Galactic LBVs included in DR1.  The distances to all four
objects are closer than traditionally assumed in the literature,
suggesting lower intrinsic luminosities.

1.  For HD~168607, the reduction in distance is about a factor of 2,
but the new distance is consistent with previous values within the
uncertainty.  We therefore postpone an evaluation of its properties
until the higher precision that will be avaialable later from \gaia.

2.  For HR~Car, the implied distance of $\sim$2.3~kpc is more than a
factor of 2 lower than the traditional value of 5$\pm$1~kpc, but the
upper limit to the distance of 4.6~kpc is still marginally consistent
with the old value.  It is interesting that at this closer distance,
HR~Car's position on the HR Diagram is nearly identical to the
progenitor of SN~1987A.

3.  The prototypical, classical LBV star AG~Carinae has a \gaia\
parallax indicating a distance of only $\sim$2~kpc---three times closer
than previously thought, making it 9 times less luminous.  The upper
limit to its distance is inconsistent with the usually adopted value
of 6.4--6.7~kpc \citep{humphreys89}.  It now lies along a track of a 
25~$M_{\odot}$ star, rather than a 90--100~$M_{\odot}$ star.  If correct, this has
dramatic implications for the interpretation of this star and its
nebula.  In particular, AG~Car may have gone through a previous RSG
phase, and it is likely to be a product of binary interaction.
Moreover, since AG~Car is regarded as a defining member of the S
Doradus class and a classical high-luminosity LBV, its lower
luminosity and mass have profound consequences for our traditional
view of LBVs in general.

4.  For Hen~3-519, the distance implied by the \gaia\ parallax is also
about 2~kpc, which is $\sim$4 times closer than previously thought
(making the star 16 times less luminous).  This moves it far away from
the traditional locus of LBVs on the HR Diagram.  Its origin and
evolutionary state remain unclear, but its nebula may be the product
of previous RSG mass loss or binary interaction.

Given that these stars were thought to be among the most
luminous stars in the Milky Way and have shaped our views of LBVs for
the last 30 years, it seems likely that \gaia\ distances for the
remaining LBVs may instigate a re-evaluation of our standard view of
LBVs.  More precise values of the parallax for these four objects and for a larger number of
Galactic LBVs will be available soon; the results reported here may
hint at a coming upheaval in our understanding of massive star evolution, even if they are regarded as preliminary.

An obvious avenue for further investigation is whether there are
similarly low-luminosity stars in the LMC or other nearby galaxies
that have LBV-like spectra or variability, but that have evaded
detection because they are faint.  On the other hand, this does not
undermine the properties of existing LBVs known in the LMC and SMC,
since their distances are not so uncertain; the S~Dor strip still
seems to work for them.  It will be interesting to investigate why
there are stars that appear to be classical S~Doradus-like LBVs in the
Milky Way that are so far below the traditional S Dor instability
strip.  Perhaps metallicity plays a key role, or perhaps samples in the LMC are incomplete at lower luminosity.

\acknowledgments

We thank Megan Kiminki for double-checking the \gaia\ distances for O-type stars in the Carina Nebula, and for helpful discussions.  Support for N.S.\ was provided by the National Science Foundation (NSF) through grants AST-1210599 and AST-1312221 to the University of Arizona.  K.G.S.\ acknowledges partial support from NSF PAARE grant AST-1358862.  This work has made use of data from the European Space Agency (ESA) mission {\it Gaia\/} (http://www.cosmos.esa.int/gaia), processed by the {\it Gaia\/} Data Processing and Analysis Consortium (DPAC, http://www.cosmos.esa.int/web/gaia/dpac/consortium). Funding for the DPAC has been provided by national institutions, in particular the institutions participating in the {\it Gaia\/} Multilateral Agreement.

\end{document}